\documentclass[12pt]{article}
\usepackage[reqno]{amsmath}
\usepackage{amssymb}
\usepackage{graphicx}

\usepackage{amsmath}
\usepackage{amssymb}

\usepackage{setspace}
\usepackage{caption}
\usepackage{cite}

\def\be{\begin{equation}}
\def\bea{\begin{eqnarray}}
\def\ee{\end{equation}}
\def\eea{\end{eqnarray}}

\newcommand{\bean}{\begin{eqnarray*}}
\newcommand{\eean}{\end{eqnarray*}}

\newcommand{\T}{\mathbb{T}}

\newcommand{\A}{&\!\!\!}

\def\be{\begin{equation}}
\def\ee{\end{equation}}
\def\bea{\begin{eqnarray}}
\def\eea{\end{eqnarray}}

\def\T11{{T}^{1,1}}
\def\bear{\begin{eqnarray}}
\def\eear{\end{eqnarray}}

\newcommand{\pa}{\partial}

\newcommand{\comment}[1]{}

\newcommand{\pasl}{\pa\kern-.55em /}

\newcommand{\ksl}{k\kern-.55em /}

\DeclareFixedFont{\xiiss}{OT1}{cmss}{m}{n}{12}
\DeclareFixedFont{\ixss}{OT1}{cmss}{m}{n}{9}
\DeclareFixedFont{\cmrnine}{OT1}{cmr}{m}{n}{9}

\newcommand{\CCs}{\hbox{\ixss C\kern-.4emI}}
\newcommand{\ZZs}{\hbox{\ixss Z\kern-.4emZ}}

\begin{document}


\vspace{18pt}
\vskip 0.01in \hfill TAUP-2688/07
\vskip 0.01in \hfill {\tt hep-th/yymmnnn}

\vspace{30pt}

\begin{center}
{\bf \LARGE Non critical holographic models of the   thermal phases of QCD}
\end{center}

\vspace{30pt}

\begin{center}
Victoria Mazo $\,^a$, and Jacob Sonnenschein$\,^{b}$,

\vspace{20pt}

\textit{School of Physics and Astronomy\\ The Raymond and Beverly Sackler
Faculty of Exact Sciences\\ Tel Aviv University, Ramat Aviv 69978,
Israel\\[10pt]
}

\end{center}


\begin{center}
\textbf{Abstract }
\end{center}
We analyze the thermal phases of a non critical  holographic model of QCD.
 The model is based on a six dimensional background of $N_c$  non extremal D4 branes
wrapping a spacial circle of  radius $R$ and the compactified Euclidean time direction
of  radius $\beta=1/T$.
 We place in this background stacks of   $N_f$ D4  and anti-D4 flavor probe branes
with a separation distance $L$  at large radial direction.
The analysis of the DBI effective action yields the following  phase diagram:
At low temperature  the system is in a confining phase with broken chiral symmetry.
In the high temperature deconfining phase chiral  symmetry can be either restored for
$L>L_c=1.06 R$ or broken for $L<L_c$.
  All of these phase transitions are of
first order.
We analyze the spectrum of the low-spin and high-spin mesons.
High spin mesons above certain critical angular momentum ``melt''.  We detect (no) drag
for ( mesons) quarks moving in hot quark-gluon fluid.
The results  resemble
the structure and properties of the thermal Sakai-Sugimoto model derived in  hep-th/0604161.

\vspace{4pt} {\small \noindent

 }  \vfill
\vskip 5.mm
 \hrule width 5.cm
\vskip 2.mm
{\small
\noindent
E-mails : $^a$victor1@post.tau.ac.il, $^b$  cobi@post.tau.ac.il
}

\thispagestyle{empty}

\eject

\setcounter{page}{1}

 \tableofcontents


\newpage{}

%
%

\section{Introduction}
Recently the phases of thermal holographic QCD (HQCD) have been analyzed
\cite{Aharony:2006da} in the context of the model of Sakai and Sugimoto \cite{SS, Sakai}.
It was found out that the confinement/deconfinement and chiral symmetry breaking/restoring
phase transitions are  first order transitions and they do  not necessarly coincide with each other.
The system may admit an intemediate deconfined phase with broken chiral symmetry.

The mesonic world at these phases was later investigated in \cite{Peeters:2006iu}.
 The temperature dependence of low-spin as well as high-spin meson masses was shown to exhibit a
pattern familiar from the lattice.
The Goldstone bosons associated with chiral symmetry breaking were shown
to disappear above the chiral symmetry restoration temperature.
 The dissociation temperature of mesons as a function of their spin was determined,
 showing that at a fixed quark mass, mesons with larger spins dissociate at lower temperatures.
It was further shown that unlike quarks,
 large-spin mesons do not experience drag effects when moving through the quark gluon fluid. They do, however, have a maximum velocity for fixed spin, beyond which they dissociate.

HQCD models    based on  critical string theories suffer from the major drawback of incorporating
undesired KK modes. Whereas the modes associated with the $S^5$ in the string theory on $AdS_5\times S^5$
are essential to describe the dual ${\cal N}=4$ SYM theory, the KK modes in models of HQCD do not correspond
to modes of the gauge theory. Moreover, the mass scale of those KK modes is the same as that of the glueballs and hadrons and there is no known method to disentangle the two scales.
The most natural way to overcome this problem is to consider strings in non-critical dimensions so as to minimize the  set of KK modes. Since the  pioneering paper   of Polyakov\cite{Polyakov}.
 there have been  many attempts to write down an non-critical string model of QCD \cite{Kuperstein:2004yk}- \cite{NCHQCD}.
The main problem with non-critical holography is the fact that the corresponding SUGRA backgrounds
have curvature of order one and there is no way to go to a region of small curvature by taking the limit of
large $\lambda_{'t Hooft}$. However, it turns out that  in fact any HQCD model even those based on critical string
theories must have eventually curvature of order one.
This is  necessary to avoid a gap in the masses
of low spin mesons holographically described by fluctuations of the flavor D-branes, and high spin mesons
\cite{Kruczenski:2004me}. There is yet another reason in favor of backgrounds with curvature which is not small.
To recast a non-trivial $a-c$ anomaly which characterizes supersymmetric QCD models, one must turn on higher curvature
terms.\cite{Schwimmer:2003eq},\cite{Alexandria}.

In this paper, we look at the non-critical $AdS_6$ black hole
solution \cite{Kuperstein:2004yk}. This model was shown
\cite{Kuperstein:2004yf} to reproduce some properties of the
4-dimensional  non-supersymmetric YM theory like an area law for
the Wilson loop, a mass gap in the glueball spectrum etc. At high
energies the theory is dual to a thermal gauge theory at the same
temperature as the black hole temperature and at low energies the
dual theory is effectively 4-dimensional pure YM.  For any
non-critical model, the curvature is of order one in units of
$\alpha'$ and hence higher order curvature correction may affect
the structure of the model. On the other hand unlike in \cite{SS}
where in what corresponds to the ``uv region'' the dilaton blows up
and one has to elevate the model to an M theory setup, the non-critical model
admits small
 small string
coupling and hence  stringy corrections can be safely ignored.

In this work we introduce flavor in this setup by adding
D4,$\overline{\textrm{D4}}$ probe branes. This is similar to the
D8,$\overline{\textrm{D8}}$ of the Sakai Sugimoto  model in critical dimension.
We analyze the classical configurations of the probe branes. At the low temperature
phase the only solution of the equations of motion is a U shape, where  the
branes and anti-branes merge together in the region that corresponds
in the dual gauge theory  to the IR. The U shape is a geometrical manifestation
of chiral symmetry breaking. At the high temperature ( deconfining )   phase
there are two possible solutions ,
again the U shape configuration and a $\mid\  \mid $ shape of parallel branes and anti-branes.
To determine the phase structure we compute the difference of the free energy
between these two configurations. The free energy is proportional to the value of
effective action. The latter  includes the DBI action and a Cern Simons term of the
form $\int c_5$ for charged probe branes and only the DBI term for charge-less ones.
It turns out that in the former case the difference of the free energies diverges
and hence it cannot correspond to the difference of the free energy between two phases
of the dual gauge theory. We therefore set this CS term to zero and perform from thereon
all the computations with out this CS term. This resembles the situation in other non-critical
models with flavor branes like \cite{Klebanov:2004ya}  and \cite {Gursoy:2007er}.

The outcome of the model is similar to that of the thermal Sakai-Sugimoto  model
\cite{Aharony:2006da}.
There are three different phases. In the \emph{low-temperature}
phase the background is the Euclidean continuation of a Lorentzian
background. Gluons are confined in this phase. After the
confinement/deconfinement transition for the gluons, there is the
\emph{intermediate-temperature} phase. In this phase gluons are
deconfined, but chiral symmetry is still broken. Mesonic bound
states still exist, as the D4-brane embedding is not yet touching
the horizon. At sufficiently high temperature, the lowest-energy
configuration of the D4-branes is the one in which they are
parallel and fall down to the horizon. This is the
\emph{high-temperature} phase, in which chiral symmetry is
restored. If the ratio $L/R > 1.06$, there is no
intermediate-temperature phase, so that the
confinement/deconfinement and the chiral symmetry breaking
transition coincide.

We also analyze the spectrum of low-spin and high-spin mesons.
Low-spin mesons correspond on the string theory side to
fluctuations of the massless fields on the probe branes. We
identify the Goldstone boson associated with the chiral symmetry
breaking. High-spin mesons, as for the critical case, can be
described as classical spinning open strings. The temperature
dependence for both low-spin and high-spin mesons is similar, that
is the masses of mesons go down as the temperature goes up. For
high-spin mesons there is a maximum value of angular momentum
beyond which mesons cannot exist and have to melt. We find the
drag force that a quark experiences moving through a hot gluon
plasma, and we also find that high-spin mesons do not experience
any drag force because for high-spin mesons at finite temperature,
one can find generalized solutions where the meson moves with
linear velocity, rigidly, with free boundary conditions in the
direction of motion. Hence one does not need to apply any force to
maintain this motion.

A main goal of this work has been to compare the phase diagram that follows
from a critical holographic model versus that one associated with a non-critical
HQCD model. Since the latter is characterized by a curvature of order one,
strictly speaking one is not allowed to ignore the higher curvature corrections
of the supergravity action. Hence there is priori no reason that the structure of the thermal
phases
that emerge from our analysis will resemble at all the one extracted from a critical
HQCD model. However, the outcome of  this paper is that in fact the results from the critical and
non-critical holographic models are very similar.

We begin in section 2 with a short review of the SS model and its
behavior at finite temperature. In section 3 we describe the
$AdS_6$ model at zero temperature. In section 4 we discuss the
behavior of this theory at finite temperature. We discuss the bulk
thermodynamics, which leads to the confinement and deconfinement
phases, and chiral symmetry restoration at a certain temperature.
In section 5 we have a close look on the spectrum of low-spin as
well as high-spin mesons in different phases and also discuss the
drag force on quarks and mesons.

%
%

\section{Review of Sakai-Sugimoto model at finite temperature}

The Sakai-Sugimoto model \cite{SS, Sakai} is based on a
D4/D8-$\overline{\text{D8}}$ brane system consisting of
 $N_c$ D4-branes compactified on S${}^1$ and $N_f$
D8-$\overline{\text{D8}}$-brane pairs transverse to the S${}^1$.
The brane configuration of the system is
\begin{equation}
\begin{tabular}{ccccccccccc}
& $t$ & $x_1$ & $x_2$ & $x_3$ & $x_4$ & $x_5$ & $\theta_1$ &
$\theta_2$ & $\theta_3$ & $\theta_4$ \\ \hline D4 & $\diamond$ &
$\diamond$ & $\diamond$ & $\diamond$ & $\diamond$ & $$
& $$ & $$ & $$ & $$ \\
D8-$\overline{\text{D8}}$ & $\diamond$ & $\diamond$ & $\diamond$ &
$\diamond$
& $$ & $\diamond$ & $\diamond$ & $\diamond$ & $\diamond$ & $\diamond$ \\
\end{tabular} \nonumber
\end{equation}
with $x_4$ and $\theta$'s being coordinates of S${}^1$ and S${}^4$
respectively.

 We look at the D4-branes in the large $N_c$ and near horizon limits. In these limits they are
classical solutions of the type IIA supergravity in ten
dimensions. This gravitational background is dual to a
five-dimensional gauge theory, which looks four-dimensional at
energy scale below the compactification scale. Imposing periodic
boundary conditions on the bosons and antiperiodic ones on the
fermions along the compactified direction, supersymmetry is
explicitly broken. The scalars and the fermions on the D4-branes
become massive and are decoupled from the system at low energy.
Thus one obtains a $U(N_c)$ pure gauge theory. To describe quarks
in the fundamental representation of the gauge group $U(N_c)$ one
introduces flavor $N_f$ $D8-\overline{D8}$ pairs into the D4
background. We assume $N_f<<N_c$ which  allows us to treat the
$N_f$ $D8-\overline{D8}$ branes as  probes.
\par
The finite temperature behavior of the Sakai-Sugimoto model was
discussed in \cite{Aharony:2006da,Peeters:2006iu,
Parnachev:2006dn,Horigome:2006xu}. As opposed to the zero
temperature case there are two solutions at finite temperature,
because one Wick-rotates the metric (generates a black hole
solution) and an asymptotic symmetry between compactified
Euclidean time coordinate (with periodicity $\beta=1/T$) and $x_4$
(with periodicity $2\pi R$) appears. In \cite{Aharony:2006da} it
was shown that one of them dominates at low temperatures and the
other one at high temperatures. A phase transition between these
backgrounds occurs at the temperature $T_c=1/2\pi R$. This phase
transition is of the first order and represents a
confinement/deconfinement transition \cite{Witten:1998zw}.

The bulk background geometry at low temperature is represented by
the following metric
\begin{eqnarray}
ds^2 \A = \A \left( \frac{u}{R_{D4}} \right)^{\frac{3}{2}}
       \left( dt^2 + \delta_{ij}dx^{i}dx^{j} + f(u) dx_4^2 \right)
       + \left( \frac{R_{D4}}{u} \right)^{\frac{3}{2}}
       \left( \frac{du^2}{f(u)} + u^2 d\Omega^2_4 \right),
       \nonumber\\
\A\A  e^{\phi} = g_s \left( \frac{u}{R_{D4}}
\right)^{\frac{3}{4}}, \qquad F_4 = dC_3 = \frac{2 \pi
N_c}{V_4}\epsilon_4, \qquad f(u) = 1 - \frac{u_\Lambda^3}{u^3},
\label{SSmetric}
\end{eqnarray}
where $d\Omega_4^2$ is the metric of S${}^4$ and $R_{D4}^3 = \pi
g_s N_c l_s^3$ with $g_s$ and $l_s$ being the string coupling and
the string length. $\epsilon_4$ and $V_4$ are the volume form and
the volume of S${}^4$. The $x_4$-$u$ submanifold has a cigar-like
form with a tip at $u=u_T$. To avoid singularity at the tip of the
cigar $x_4$ should be periodic with periodicity
\begin{equation}
\delta
x_4=\frac{4\pi}{3}\left(\frac{R^3_{D4}}{u_\Lambda}\right)^{1/2}=2\pi R
\end{equation}
 The effective action of the D8-branes consists of the DBI action
and the Chern-Simons term
\begin{equation}
S_{\text{D8}} = T_8 \int d^9x \, e^{-\phi} \, \text{Tr} \sqrt{\det
(g_{MN} + 2\pi\alpha'F_{MN})} - \frac{i}{48 \pi^3}
\int_{\text{D8}}C_3 \; \text{Tr} F^3, \label{dbics}
\end{equation}
where $g_{MN}$ and $F_{MN}$ are the induced metric and the field
strength on the D8-brane and $T_8$ is the tension of the D8-brane.
The CS term in the D8-action does not affect the solution of the
equation of motion of the gauge field since it has a classical
solution of a vanishing gauge field.

\begin{figure}[t]
\begin{center}
\scalebox{1.0}{\includegraphics{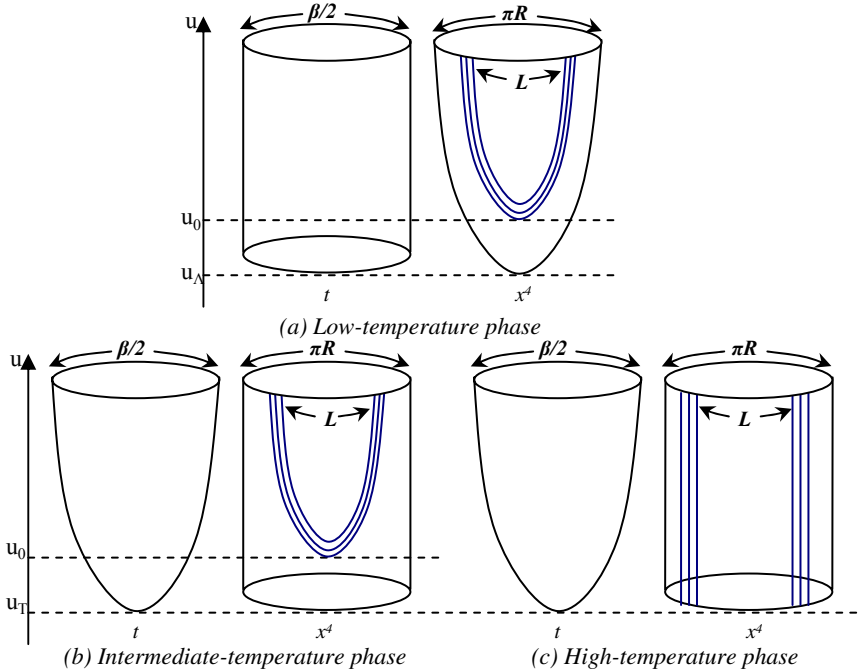}} \caption{The
configuration of flavor branes at three phases: (a)
low-temperature phase, (b) intermediate and (c) high-temperature
phases.\label{phasediag}}
\end{center}
\end{figure}

The Hamiltonian of the action does not depend on $x_4$  and therefore the equation of
motion equals to a constant. To solve it we assume that there is a
point $u_0$ where the profile $u(x_4)$ has a minimum
($u'|_{u=u_0}=0$). The form of the profile is drawn in figure
\ref{phasediag}(a).

The $x_4$ circle shrinks to zero at $u=u_\Lambda$. Therefore D8 branes
and antibranes have no place to end and should stay all the time
connected. Because of this configuration the chiral symmetry
$U(N_f)_L\times U(N_f)_R$ on the probe D8-D8 pairs is always
broken to a diagonal subgroup $U(N_f)_V$ in the low temperature
phase.

In the high temperature phase the preferred background is the one
with the interchanged role of the $t$ and $x_4$ circles (by moving
the factor of $f(u)$ in (\ref{SSmetric}) from the $dx_4^2$ term to
the $dt^2$ term). Now the $t$-circle shrinks to zero at $u_T$
(which is now related to $T$ rather that to $R$), while the $x_4$
circle never shrinks. While the configuration with connected
flavor branes is still possible, a new configuration with parallel
branes appears (see figure \ref{phasediag}(b) and
\ref{phasediag}(c)). This new configuration is also a solution of
the equation of motion of the new background DBI action and it
means that chiral symmetry $U(N_f)_L\times U(N_f)_R$ is restored.

Both configuration are possible at the high temperature phase
(deconfinement phase). To see when they are preferred we need to
compute their free energy (the one with the lower free energy is
preferred in the given temperature range). In
\cite{Aharony:2006da} it was found that in the range $T_c\leq T <
T_{\chi SB}$ ($T_{\chi SB}=0.154/L$, $L$ is the separation
distance between the flavor branes at $u\rightarrow\infty$) the
preferred configuration is with connected branes and at $T\geq
T_{\chi SB}$ - with parallel branes. Therefore deconfinement and
chiral symmetry restoration do not occur together but there is an
intermediate phase with deconfinement and broken chiral symmetry.


%
%

\section{Near extremal $AdS_6$ model with flavor branes at zero temperature}

We are interested in the non-critical flavored version of the model of
\cite{SS}, which was considered in
\cite{Kuperstein:2004yk},\cite{Casero:2005se}. The starting point
is to consider unflavored conformal $AdS_6$ background, which is
the dual of a fixed point non-supersymmetric 5-dimensional gauge
theory without fundamental quarks. The construction of the model
can be made by either first taking the near extremal limit of the
$AdS_6$ background and then adding flavors, or by adding flavors
first and then taking the near extremal limit of the flavored
$AdS_6$. We follow the former one. The non-critical version of the
near horizon limit of $N_c$ near extremal $D4$-branes wrapped over
a circle with anti-periodic boundary conditions takes the form of
a static black hole embedded inside $AdS_6$. The only surviving
fermionic degrees of freedom are excited Kaluza-Klein modes
because the anti-periodic boundary conditions project massless
fermions out of the spectrum. At high energies the system is dual
to a thermal gauge theory at the same temperature as the black
hole temperature and at low energies the KK modes can not be
excited and the dual theory is effectively 4-dimensional pure YM.

The 6-dimensional background metric, 6-form field strength and
constant dilaton are given by
\begin{align}\label{ds6}
&
ds^2_6=\bigg(\frac{u}{R_{AdS}}\bigg)^2(-dt^2+\delta_{ij}dx^idx^j+f(u)dx_4^2)+
\bigg(\frac{R_{AdS}}{u}\bigg)^2\frac{du^2}{f(u)}\nonumber\\
& F_{(6)}=Q_c\bigg(\frac{u}{R_{AdS}}\bigg)^4dx_0\wedge dx_1\wedge
dx_2\wedge dx_3\wedge du\wedge dx_4\nonumber\\
& e^\phi=\frac{2\sqrt{2}}{\sqrt{3}Q_c} \quad \quad
R_{AdS}^2=\frac{15}{2} \quad \quad
f(u)=1-\bigg(\frac{u_{\Lambda}}{u}\bigg)^5
\end{align}
The space spanned by $u$ and $x_4$ has a topology of a cigar with
the minimum value $u_{\Lambda}$ at its tip. To avoid a conical
singularity at the origin, $x_4$ needs to be periodic with
periodicity
\begin{equation} \label{x4period}
x_4\sim x_4+\frac{4\pi R_{AdS}^2}{5u_{\Lambda}}=x_4+2\pi R
\end{equation}
The typical mass scale below which the theory is effectively
4-dimensional is
\begin{equation}
M_{\Lambda}=\frac{2\pi}{\delta
x_4}=\frac{5}{2}\frac{u_{\Lambda}}{R^2_{AdS}}
\end{equation}

Since the gauge theory is not supersymmetric there are two
ways to add flavor to the $AdS_6$ black hole background  - by
adding D4- or D5-probe branes. But it seems natural to include
probe $D4$-branes and antibranes extended along the Minkowski
directions and stretching to infinity in the radial direction
since then the low energy limit of the gauge theory will contain
massless fundamental quarks, while adding D5-branes, which need to
wrap the $S^1$, due to the antiperiodic boundary conditions on
$S^1$ will generate mass to the quarks of the 4-dimensional gauge
theory. When all the quarks are massless one can
reproduce a spontaneous chiral symmetry breaking in terms of the
string dual theory. The brane configuration looks the following
way
\begin{equation}
\begin{tabular}{ccccccc}
& $t$ & $x_1$ & $x_2$ & $x_3$ & $x_4$ & $x_5$ \\
\hline D4 & $\diamond$ & $\diamond$ & $\diamond$ & $\diamond$ &
$\diamond$ &
$$ \\
D4-$\overline{\text{D4}}$ & $\diamond$ & $\diamond$ & $\diamond$ &
$\diamond$ & $$ & $\diamond$  \\
\end{tabular} \nonumber
\end{equation}
In the limit of large $N_c$ and very small $g_s$ (with fixed
$g_sN_c$), $N_f\ll N_c$ and $L\gg l_s$, the coupling of the
strings stretching between two D4-probe branes, two D4-probe
antibranes or between a D4-probe brane and an antibrane goes to
zero and they become non-dynamical sources. Hence, the degrees of
freedom in the low energy limit and in the above limiting case are
described by the strings stretching between color branes or
between a color brane and a probe brane/antibrane. The gauge
symmetry of the flavor branes $U(N_f)\times U(N_f)$ becomes a
global symmetry of QCD and represents a chiral symmetry of the
quarks. The fermions that appear from the color - D4-probe branes
intersection transform as $(\bar{N_f},1)$ of the global symmetry
and that from the color - $\overline{D4}$-probe branes
intersection transform as $(1,\bar{N_f})$. Both fermions transform
in the fundamental $N_c$ representation of the color group.

The picture is similar to the Sakai-Sugimoto model
\cite{Sakai,Aharony:2006da} but the background we consider is
non-critical. For any non-critical model, the curvature is of
order one in units of $\alpha'$. But taking large $N_c$ limit
guarantees small string coupling and one expects that stringy
corrections will not affect calculations on the non-critical
gravity side. The results in \cite{Kuperstein:2004yf} are at least
of the same order of magnitude as those given by experiments or
lattice calculations, showing therefore that this assumption is
not meaningless.

We consider the action
\begin{align} \label{S_D4}
S_{D4}=T_4\int
d^5xe^{-\phi}\sqrt{-det\hat{g}}-\tilde{a}T_4\int\mathcal{P}(
C_{(5)})
\end{align}
where $\hat{g}$ is the induced metric over D4-brane worldvolume
and $\mathcal{P}( C_{(5)})$ is the pull-back of the RR 5-form
potential over the D4-brane worldvolume. Taking $\tilde{a}$ is a
constant which fixes the relative strength of the DBI and CS terms
in (\ref{S_D4}). $\tilde{a}$ should be taken equal to zero if WZ
coupling is not present at all. $\tilde{a}$ equals to one would be
the direct naive generalization from the 10-dimensional theory,
but it is not well understood what should be written in this
two-derivative approximation to the non-critical setup.

 The induced metric on the D4-branes
is
\begin{align}
ds^2_6=\bigg(\frac{u}{R_{AdS}}\bigg)^2(-dt^2+\delta_{ij}dx^idx^j)+
\bigg(\frac{u}{R_{AdS}}\bigg)^2\bigg(f(u)+\bigg(\frac{R_{AdS}}{u}\bigg)^4\frac{u'^2}{f(u)}\bigg)dx_4^2
\end{align}
Substituting the determinant of the metric and the pullback of the
RR 5-form potential $C_{(5)}$ into the action (\ref{S_D4}),we get
\begin{align} \label{S_D4'}
S_{D4}=\hat{T}_4e^{-\phi}\int
dx_4\bigg(\frac{u}{R_{AdS}}\bigg)^5\bigg[\sqrt{f(u)+
\bigg(\frac{R_{AdS}}{u}\bigg)^4\frac{u'^2}{f(u)}}-a\bigg]
\end{align}
where $\hat{T}_4$ includes the outcome integration over all
coordinates apart from $dx_4$ and $a\equiv
\frac{2}{\sqrt{5}}\tilde{a}$.

The action does not depends explicitly on $x_4$ therefore the
Hamiltonian will be conserved:
\begin{equation}\label{}
  \bigg(\frac{u}{R_{AdS}}\bigg)^5 \bigg(\frac{f(u)}{\sqrt{f(u)+\big(\frac{R_{AdS}}{u}\big)^4\frac{u'^2}{f(u)}}}-a\bigg)=
  \bigg(\frac{u_0}{R_{AdS}}\bigg)^5(\sqrt{f(u_0)}-a)
\end{equation}
where $u_0$ is a point of a vanishing profile
$u'(u)\big|_{u_0}=0$.

Defining $y\equiv\frac{u}{u_0}$,
$y_{\Lambda}=\frac{u_{\Lambda}}{u_0}$, $f(y)\equiv
1-\big(\frac{y_{\Lambda}}{y}\big)^5$, the profile reads
\begin{equation}\label{u'}
u'=\bigg(\frac{u}{R_{AdS}}\bigg)^2f(y)\sqrt{\frac{f(y)}{(y^{-5}\sqrt{f(1)}+a(1-y^{-5}))^2}-1}
\end{equation}
At $u\rightarrow\infty$ we want $N_f$ D4-branes to be localized at
$x_4=0$ and $N_f$ $\overline{D4}$-branes at $x_4=L$. These branes
can't go to the interior of the space because they don't have
where to end inside the "cigar". Therefore they should smoothly
connect at some point $u=u_0$ ($u_0\leq u_{\Lambda}$) and
therefore at zero temperature chiral symmetry is broken.

We can express $L$ as a function of $u_0$, $u_{\Lambda}$ and
$R_{AdS}$:
\begin{align}\label{}
 L&=\int_0^L dx_4=2\int_{u_0}^{\infty}\frac{du}{u'}\nonumber\\
 &=2u_0\int_1^{\infty}dy\bigg(\frac{R_{AdS}}{u}\bigg)^2
 \frac{1}{f(y)}\frac{1}{\sqrt{\frac{f(y)}{(y^{-5}\sqrt{f(1)}+a(1-y^{-5}))^2}-1}}
\end{align}
Setting $z\equiv y^{-5}$, we get
\begin{equation}\label{}
L=\frac{2R_{AdS}^2}{5u_0}\int_0^1dz\frac{1}{z^{\frac{4}{5}}(1-y^5_{\Lambda}z)}
\frac{z(1-y^5_{\Lambda})+a(1-z)}
{\sqrt{1-y^5_{\Lambda}z-(z(1-y^5_{\Lambda})+a(1-z))^2}}
\end{equation}
From here we see that small values of $L$ correspond to large
values of $u_0$ and to $y_{\Lambda}<<1$. In this limit
$L\varpropto\frac{R_{AdS}^2}{u_0}$. The general dependence of $L$
on $u_0$ is more complicated.

%
%

\section{Near extremal $AdS_6$ model with flavor $D4$-$\overline{D4}$ branes
 at finite temperature}

\subsection{Bulk thermodynamics}

We consider flavor branes as probes with $N_f\ll N_c$ and
therefore we can analyze the  thermodynamics of our model at
finite temperature by considering only background geometry and
then add probe D4-branes to the dominant bulk background at each
temperature. In the gravity approximation (large $N_c$ limit) we
should look at Euclidean backgrounds, which are asymptotically
(\ref{ds6}), but with Euclidean and periodic time with a
periodicity $t=1/T=\beta$ and with anti-periodic boundary
conditions for the fermions along the time direction in addition
to the $x_4$ direction. This background is just a Euclidian
continuation of the background (\ref{ds6}) with the $x_4$ compact
direction with a periodicity $2\pi R$ (with $R$ related to
$u_{\Lambda}$ by (\ref{x4period})) that shrinks to zero at $u=u_0$
and with the time direction that remains always finite with an
arbitrary periodicity equal to $\beta$ (see figure
\ref{phasediag}(a)).

But now we can consider another solution with the same
asymptotics, which is given by exchanging the behavior of the $t$
and $x_4$ circles (i.e. by moving $f(u)$ in the metric (\ref{ds6})
from the $dx_4^2$ term to the $dt^2$ term). Then now the time
direction shrinks to zero size at $u=u_T$ ($u_T$ is related to
$\beta$), while the $x_4$ circle never shrinks (see figure
\ref{phasediag}(b)).

\subsection{The bulk free energies of the low and high
temperature phases}

In order to decide which background dominates at a given
temperature we need to compute their free energies. We look at the
difference between the free energies, which is proportional to the
difference between the actions of the backgrounds times the
temperature (in the gravitational approximation), because it turns
out to be finite despite that classical actions might diverge. In
our calculations we use the notations and the results for the
action computed in \cite{Kuperstein:2004yk}.

The class of Euclidean metrics that we are looking on can be
parameterized as
\begin{equation}\label{}
   l_s^{-2}ds^2=d\tau^2+e^{2\lambda(\tau)}dx^2_{||}+e^{2\tilde{\lambda}(\tau)}dx^2_c
\end{equation}
where $x_c$ is either $x_4$ or $t$ (the one whose circle shrinks
to zero size at the minimal value of $u$ at a certain
temperature), $x_{||}$ are the other four coordinates of $R^{4,1}$
(one of which is also compactified), $\tau$ is the radial
direction and
\begin{align}\label{variables}
 e^{2\lambda}=\bigg(\frac{u}{R_{AdS}}\bigg)^2 \quad \quad
 e^{2\tilde{\lambda}}=\bigg(\frac{u}{R_{AdS}}\bigg)^2 \bigg(1-\bigg(\frac{u_{\Lambda}}{u}\bigg)^5\bigg)
\end{align}
The functions $\lambda$ and $\tilde{\lambda}$ depend only on the
radial coordinate. The color D4-brane wrap the circle $x_c$, while
the flavor D4-brane are points on the circle. The background also
includes a constant dilaton $\phi_0$ and a 6-form RR field
strength. We define a deformed dilaton $\varphi$ and a new radial
coordinate $\rho$ as
\begin{align}\label{}
  & \varphi=2\phi_0-4\lambda-\tilde{\lambda}\nonumber\\
  & d\tau=-e^{-\varphi}d\rho
\end{align}
Since the background depends only on a radial direction, the sugra
action reduces to the following (0+1)-dimensional action:
\begin{align}\label{S_bulk}
    S&=V\int d\rho \big(-4(\lambda')^2-(\tilde{\lambda}')^2+
    (\varphi')^2+4e^{-2\varphi}-Q_c^2e^{4\lambda+\tilde{\lambda}-\varphi}
    \big)\nonumber\\
&  =-V\int_{u_{\Lambda}}^\infty du \bigg[
\big(-4\dot{\lambda}^2-\dot{\tilde{\lambda}}^2+
    \dot{\varphi}^2\big)\frac{du}{d\rho}+\bigg(4e^{-4\phi_0}
    -Q_c^2e^{-2\phi_0}\bigg)e^{8\lambda+2\tilde{\lambda}}
    \frac{d\rho}{du}\bigg]
\end{align}
where $V$ is the volume of all other directions except $\rho$ in
string units and $Q_c$ is a constant that corresponds to the
contribution of the RR flux. After rewriting the action in terms
of integrals over $u$ the minus sign arises because $d\rho/du$ is
negative (dots denote derivatives with respect to $u$). Here we
wrote the solution of the low temperature phase, for the high
temperature phase one should replace $u_\Lambda$ with $u_T$.

The equations of motion associated with the action (\ref{S_bulk})
are
\begin{align}\label{eom}
&
\lambda''-\frac{1}{2}Q_c^2e^{2(4\lambda+\tilde{\lambda}-\phi_0)}=0\\
\nonumber &
\tilde{\lambda}''-\frac{1}{2}Q_c^2e^{2(4\lambda+\tilde{\lambda}-\phi_0)}=0
\end{align}
The most general solution of this system is
(\cite{Kuperstein:2004yk}):
\begin{align}\label{sol_eom}
&
\lambda=-\frac{1}{5}\ln(\sinh(-5b\rho))+4b\rho\\
\nonumber & \tilde{\lambda}=-\frac{1}{5}\ln(\sinh(-5b\rho))-b\rho
\end{align}
where $b=-\frac{1}{\sqrt{10}}Q_ce^{-\phi_0}$.

Substituting (\ref{variables}) into (\ref{S_bulk}) we get
\begin{align}\label{S}
    S&  =V\int_{u_{\Lambda}}^\infty du \bigg[
\bigg(\frac{20}{u^2}\frac{1}{1-\big(\frac{u_{\Lambda}}{u}\big)^5}\bigg)\frac{du}{d\rho}\nonumber\\
&+
\bigg\{\big(4e^{-4\phi_0}-Q_c^2e^{-2\phi_0}\big)\bigg(\frac{u}{R_{AdS}}\bigg)^{10}
\bigg( 1-\bigg( \frac{u_{\Lambda}}{u} \bigg)^5 \bigg)\bigg\}
\frac{d\rho}{du}\bigg]
\end{align}
Using the solutions of the equations of motion associated with the
above action (\ref{sol_eom}) and (\ref{variables})
\begin{align}\label{}
& \tilde{\lambda}-\lambda=5b\rho\\ \nonumber
  &
  \frac{e^{2\tilde{\lambda}}}{e^{2\lambda}}=
  1-\bigg(\frac{u_{\Lambda}}{u}\bigg)^5=e^{10b\rho}\nonumber
\end{align}
we find that
\begin{equation}\label{}
\frac{d\rho}{du}=\frac{1}{2bu}\bigg(\frac{u_{\Lambda}}{u}\bigg)^5\frac{1}{1-\big(\frac{u_{\Lambda}}{u}\big)^5}
\end{equation}
Substituting the expression into the action we find that the
divergence at large $u$ is independent of $u_{\Lambda}$, so it
makes sense to subtract the expressions with $u_{\Lambda}$ and
with $u_T$ to obtain a finite answer. The result for the
difference between the action densities in the low temperature
phase and in the high temperature phase is given by (defining
$\hat{b}=b/u_{\Lambda}^5$ which is constant independent of
$u_{\Lambda}$)
\begin{align}\label{S}
\frac{\Delta S}{V_3}\equiv\frac{S_{low}-S_{high}}{V_3}= \frac{2\pi
R\beta}{l_s^2}\bigg(2\hat{b}+\big(4e^{-2\phi_0}-
Q_c^2\big)e^{-2\phi_0}\frac{1}{10\hat{b}R_{AdS}^{10}}\bigg)(u^5_{T}-u^5_{\Lambda})
\end{align}
Using (\ref{x4period}) and (\ref{beta}) we find
\begin{align}
u_{\Lambda}=\frac{2 R_{AdS}^2}{5R} \quad \textrm{and}\quad
u_{T}=\frac{4\pi R_{AdS}^2}{5\beta}
\end{align}
Therefore the action is proportional to
\begin{align}\label{S}
\Delta S  \propto
N_c^2\left(\frac{1}{(\beta/2\pi)^5}-\frac{1}{R^5}\right)
\end{align}
This result should be compared to the result derive in the critical
model\cite{Aharony:2006da}  In that case the power of $\beta$ and $R$ in the denominators were
found to be six. A dependence on the fifth power as we found here seems to be
more adequate for a dual five dimensional field theory.
Both backgrounds have equal free energy when both circles are
equal, i.e. $\beta=2\pi R$. When the temperature is less than
$T_d=1/2\pi R$ the background in which $x_4$ circle shrinks to
zero size dominates and when temperature is greater than
$T_d=1/2\pi R$ the background with $t$ circle shrinking to zero
dominates. There is a phase transition of first order here since
two different configurations are possible at the transition point.
If we compute the quark-antiquark potential (using the methods of
\cite{Maldacena:1998im,Rey:1998ik,Kinar:1998vq}) , which is
proportional to $\sqrt{g_{tt}g_{xx}}$, in the two backgrounds, we
find that in the low-temperature background it is finite at $u_0$
and linear, corresponding to a confined phase, and in the
high-temperature it decays, corresponding to a deconfined phase.


\subsection{Low temperature phase}

As described above the background corresponding to the low
temperature phase is the one with the $x_4$ circle shrinking to
zero at $u=u_0$. The only difference from the zero temperature
case is that the time direction is Euclidean and compactified with a
circumference $\beta=1/T$. Hence, at low temperatures the dual
gauge theory is in the confining phase. When we add flavor branes
and anti-brains to the background they have no other possibility
than to connect because $x_4$ shrinks to zero size. Therefore
chiral symmetry is broken at least until the temperatures
corresponding to deconfinement.

The metric is
\begin{align}
&
ds^2_6=\bigg(\frac{u}{R_{AdS}}\bigg)^2(dt^2+\delta_{ij}dx^idx^j+f(u)dx_4^2)+
\bigg(\frac{R_{AdS}}{u}\bigg)^2\frac{du^2}{f(u)}\nonumber\\
& f(u)=1-\bigg(\frac{u_{\Lambda}}{u}\bigg)^5 \nonumber\\
&   x_4\sim x_4+2\pi R=x_4+\frac{4\pi R^2_{AdS}}{5u_{\Lambda}}
\quad \textrm{and} \quad t\sim t+\beta \textrm{    ($\beta$
arbitrary)}
\end{align}
In the next section we will see that we can make sense of a holographic duality
only for the case of chargeless branes with vanishing $\tilde a$
Setting $\tilde{a}$ to zero, we find from (\ref{u'})
\begin{equation}\label{u'a0}
u'=\bigg(\frac{u}{R_{AdS}}\bigg)^2f(y)\sqrt{y^{10}\frac{f(y)}{f(1)}-1}
\end{equation}
Substituting (\ref{u'a0}) into (\ref{S_D4}) and using the
definitions $y\equiv\frac{u}{u_0}$,
$y_{\Lambda}=\frac{u_{\Lambda}}{u_0}$, $f(y)\equiv
1-\big(\frac{y_{\Lambda}}{y}\big)^5$ and $z\equiv y^{-5}$, we get
the following DBI action:
\begin{align}\label{}
S_{DBI} & =\hat{T}_4e^{-\phi}\int
dx_4\bigg(\frac{u}{R_{AdS}}\bigg)^5\sqrt{f(u)+
\bigg(\frac{R_{AdS}}{u}\bigg)^4\frac{u'^2}{f(u)}}\nonumber\\
& =\frac{2\hat{T_4}e^{-\phi}u_0^4}{5R_{AdS}^3} \int_0^1
dz\frac{1}{z^{\frac{9}{5}}} \frac{1}
{\sqrt{1-y_{\Lambda}^5z-z^2(1-y_{\Lambda}^5)}}
\end{align}

The relation between $L$ and $u_0$ is the same as at zero
temperatures. For small $L$ the dependence of the action on $L$
is:
\begin{equation}\label{}
  S_{DBI}\propto\frac{\hat{T_4}e^{-\phi}}{R_{AdS}^3L^4}
\end{equation}

\subsection{Intermediate and high temperature phases}

In the high temperature phase the background metric takes the form
\begin{align}\label{ds6high}
&
ds^2_6=\bigg(\frac{u}{R_{AdS}}\bigg)^2(f(u)dt^2+\delta_{ij}dx^idx^j+dx_4^2)+
\bigg(\frac{R_{AdS}}{u}\bigg)^2\frac{du^2}{f(u)}\nonumber\\
& f(u)=1-\bigg(\frac{u_{T}}{u}\bigg)^5
\end{align}
Now the time circle shrinks to zero at the minimal value of
$u=u_T$ and to avoid a singularity there the time direction should
be identified with the periodicity
\begin{equation}\label{beta}
t\sim t+\beta=t+\frac{4\pi R_{AdS}^2}{5u_T}
\end{equation}
On the other hand the periodicity of $x_4$ is now arbitrary:
\begin{equation}
x_4\sim x_4+2\pi R
\end{equation}
D4-branes span the same coordinates as previously and are
described by some profile $u(x_4)$.  The induced metric and the
DBI action now takes the form
\begin{align}
ds^2_6=\bigg(\frac{u}{R_{AdS}}\bigg)^2(f(u)dt^2+\delta_{ij}dx^idx^j)+
\bigg(\frac{u}{R_{AdS}}\bigg)^2\bigg(1+\bigg(\frac{R_{AdS}}{u}\bigg)^4\frac{u'^2}{f(u)}\bigg)dx_4^2
\end{align}
We get the following action:
\begin{align} \label{S_D4high}
S_{D4}&=T_4\int
d^5xe^{-\phi}\sqrt{-det\hat{g}}-\frac{\sqrt{5}a}{2}T_4\int\mathcal{P}(
C_{(5)})\nonumber\\
& =\hat{T}_4e^{-\phi}\int
dx_4\bigg(\frac{u}{R_{AdS}}\bigg)^5\bigg[\sqrt{f(u)}\sqrt{1+
\bigg(\frac{R_{AdS}}{u}\bigg)^4\frac{u'^2}{f(u)}}-a\bigg]\nonumber\\
&=2\hat{T}_4e^{-\phi}\bigg[\int_{u_0}^{\infty}
\frac{du}{u'}\bigg(\frac{u}{R_{AdS}}\bigg)^5\sqrt{f(u)}\sqrt{1+
\bigg(\frac{R_{AdS}}{u}\bigg)^4\frac{u'^2}{f(u)}}\nonumber\\
&\quad-a\int_{u_0}^{\infty}
\frac{du}{u'}\bigg(\frac{u}{R_{AdS}}\bigg)^5\bigg]
\end{align}
where here we turned on the CS term and took non-vanishing $a\leq 1$.
Conservation of the Hamiltonian of (\ref{S_D4high}) implies that
\begin{equation}\label{Hamilt}
 \left(\frac{u}{R_{AdS}}\right)^5\bigg( \frac{\sqrt{f(u)}}
 {\sqrt{1+\big(\frac{R_{AdS}}{u}\big)^4\frac{u'^2}{f(u)}}}-a\bigg)=
const
\end{equation}

\begin{figure}[t]
\begin{center}
\scalebox{1.0}{\includegraphics{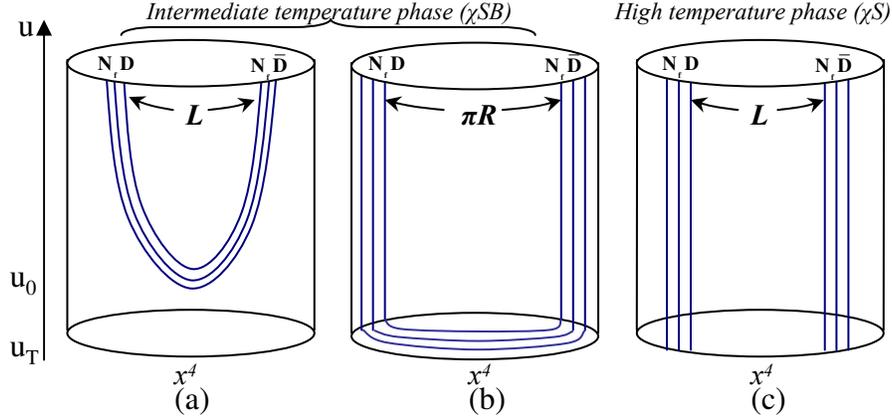}} \caption{The three
possible configurations of the flavor D4-branes and antibranes at
intermediate and high temperatures: (a) a configuration of
connected branes at the minimum $u=u_0$ with an asymptotic
separation $L$ at $u \rightarrow\infty$ (intermediate temperature
phase), (b) a configuration of connected branes in the case
$u_0=u_T$ (intermediate temperature phase), (c) parallel branes
configuration (high temperature phase).\label{phasediag'}}
\end{center}
\end{figure}

There is a solution for a vanishing profile at some point $u_0\geq
u_T$, where  $u'(u)\big|_{u_0}=0$ (see figures
\ref{phasediag'}(a) and \ref{phasediag'}(b)). This is a solution
for branes and anti-branes connected at $u=u_0$,
where  $u'$ should be zero. At low temperatures this was the
only possible configuration,
but since in the high temperature phase the $x_4$ circle never
shrinks to zero size we can consider a configuration of
non-intersecting branes and antibranes that end on the horizon of
the black hole (see figure \ref{phasediag'}(c)). The branes and
antibranes stay disconnected in the $u-x_4$ submanifold with
constant values $x_4(u)=0,L$, e.g chiral symmetry is restored.
Since the branes are now parallel to each other
 $u'=\infty$. Eventhough this is not   a solution of (\ref{Hamilt})
it is a solution of the equation of motion  associated with the action
(\ref{S_D4high})
  \footnote{We thank O. Aharoni for pointing
this to us}.


\begin{figure}[t]
\begin{center}
\includegraphics[width=.47\textwidth]{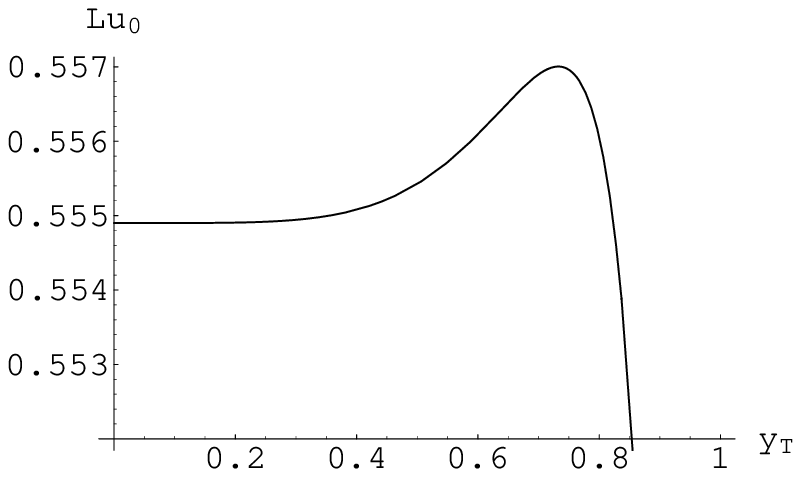}\quad
\includegraphics[width=.47\textwidth]{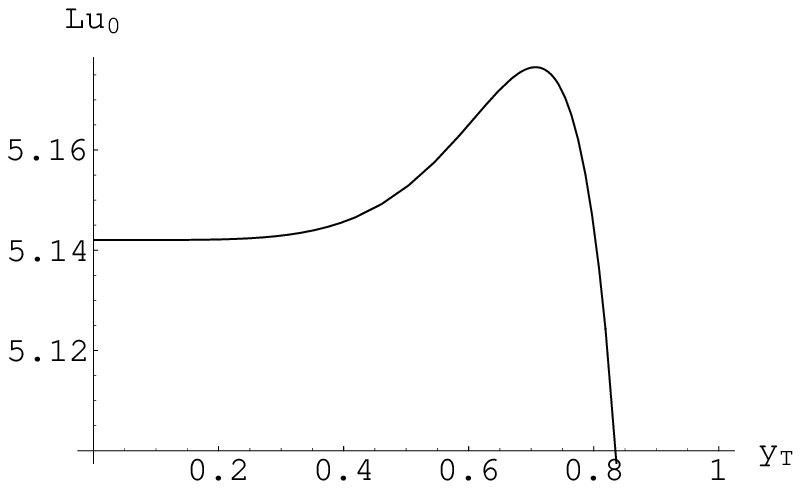}
\end{center}
\caption{$Lu_0$ as a function of $y_T$. The left graph is for the
case $\tilde{a}=0$ and the right one is for the case
$\tilde{a}=1$.\label{Lu}}
\end{figure}

Inspite of the fact that the both the parallel and the U shape solutions exist
for non trivial $a$ we found that for that case ( as is desribed in the appendix)
the difference of the free energies diverges. Thus we switch off again the CS term.
Defining $y\equiv\frac{u}{u_0}$, $y_T=\frac{u_T}{u_0}$,
$f(y)\equiv 1-\big(\frac{y_T}{y}\big)^5$, the profile velocity now
reads
\begin{equation}\label{u_tag_high}
 u'=\bigg(\frac{u}{R_{AdS}}\bigg)^2\sqrt{f(y)}\sqrt{y^{10}\frac{f(y)}{f(1)}-1}
\end{equation}

\begin{figure}[t]
\begin{center}
\vspace{3ex}
\scalebox{.8}{\includegraphics[width=.65\textwidth]{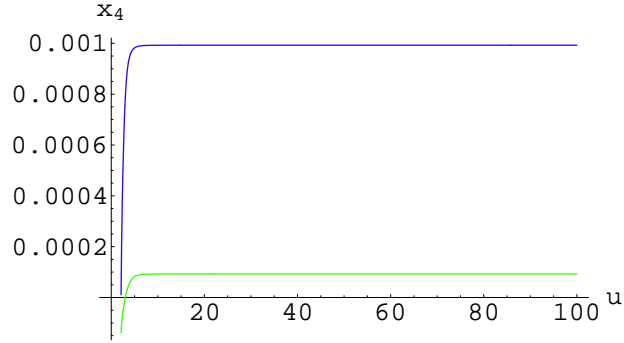}}
\end{center}
\caption{Profile $x_4$ as a function of u. The lower curve (green)
is for $u_T=2$, $u_0=3$. The upper curve (blue) is for the case
$u_0\thickapprox u_T=2$ ($R_{AdS}=1$).
 \label{profileAdS6}}
\end{figure}

Then $x_4$ as a function of $u$ becomes
\begin{align}\label{S_DBIhigh}
x_4(u)=\int_{u_0}^{u}d\hat{u}\frac{1}{\hat{u}'}=\int_{u_0}^{u}d\hat{u}\frac{1}
{\big(\frac{\hat{u}}{R_{AdS}}\big)^2\sqrt{1-(\frac{u_T}{\hat{u}})^5}
\sqrt{\big(\frac{\hat{u}}{u_0}\big)^5\frac{1-(\frac{u_T}{\hat{u}})^5}{1-u_T^5}-1}}
\end{align}
It is shown in figure {\ref{profileAdS6}}. At the beginning when
$u\approx u_0$ the profile of the branes growth very rapidly and
when $u\rightarrow \infty$ the profile is almost straight. For
$u_0\thickapprox u_T$ the profile is drawn at figure
\ref{phasediag'}(b).

\begin{figure}[t]
\begin{center}
\vspace{3ex}
\includegraphics[width=.65\textwidth]{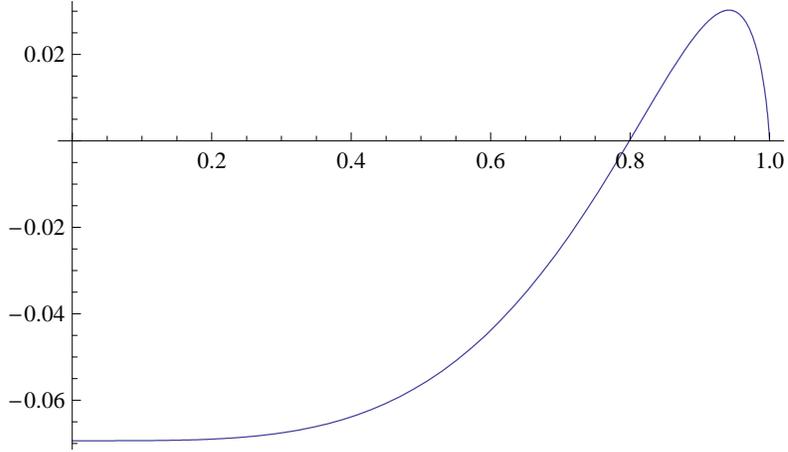}
\end{center}
\caption{ $\Delta S$ as a function of $y_T$, in units of
$2\hat{T_4}e^{-\phi}u_0^4/R_{AdS}^3$ in the case $\tilde{a}=0$.
\label{deltas}}
\end{figure}

By substituting the outcome of the equation of motion into the
action, we get
\begin{align}\label{S_DBIhigh}
S_{DBI}^{high}&=\frac{2\hat{T_4}e^{-\phi}u_0^4}{R_{AdS}^3}\int_1^{\infty}dy
\frac{y^3}{\sqrt{1-\frac{f(1)}{f(y)y^{10}}}}
\end{align}
The velocity of the profile of a parallel branes configuration is
always $u'\rightarrow\infty$ and therefore the action reads
\begin{align}\label{S_DBIhigh}
{S_{DBI}^{high}}_{u'\rightarrow\infty}&=
2\hat{T_4}e^{-\phi}\int_{u_{T}}^{\infty}du\bigg(\frac{u}{R_{AdS}}\bigg)^5
\bigg(\frac{R_{AdS}}{u}\bigg)^2\nonumber\\
&=\frac{2\hat{T_4}e^{-\phi}u_0^4}{R_{AdS}^3}\bigg[\int_1^{\infty}dy
y^3+\int_{y_T}^1dy y^3\bigg]
\end{align}
To find whether a configuration with $\chi$SB or with a restored
$\chi$S is preferred we can compute the difference between the
actions of the two configuration that is proportional to free
energy. Configuration that has a lower free energy is preferred.
\begin{align}\label{deltaS}
\Delta S&\equiv\frac{R_{AdS}^3}{2\hat{T_4}e^{-\phi}u_0^4}
(S_{DBI}^{high}-{S_{DBI}^{high}}_{u'\rightarrow\infty})\nonumber\\
&=\int_1^{\infty}dy y^3
\bigg[\frac{1}{{\sqrt{1-\frac{f(1)}{f(y)y^{10}}}}}-1\bigg]
-\int_{y_T}^1dy y^3\nonumber\\
&=\frac{1}{5}\int_0^1dz
\frac{1}{z^{\frac{9}{5}}}\bigg[\sqrt{\frac{1-y_T^5z}
{1-y_T^5z-z^2(1-y_T^5)}}-1\bigg]-\frac{1}{4}(1-y_T^4)
\end{align}
where was introduced $z=y^{-5}$ change of variables. $\Delta S$ as
a function of $y_T$ is drawn in figure \ref{deltas}. When
$y_T>0.8$ $\Delta S$ is positive, i.e.
${{S_{DBI}^{high}}_{u'\rightarrow\infty}}$ has a lower free energy
and is preferred. In this phase D4-brane are disconnected and
chiral symmetry is restored, while when $0<y_T<0.8$ D4-branes are
smoothly connected and chiral symmetry is broken.

\begin{figure}[t]
\begin{center}
\scalebox{1.0}{\includegraphics{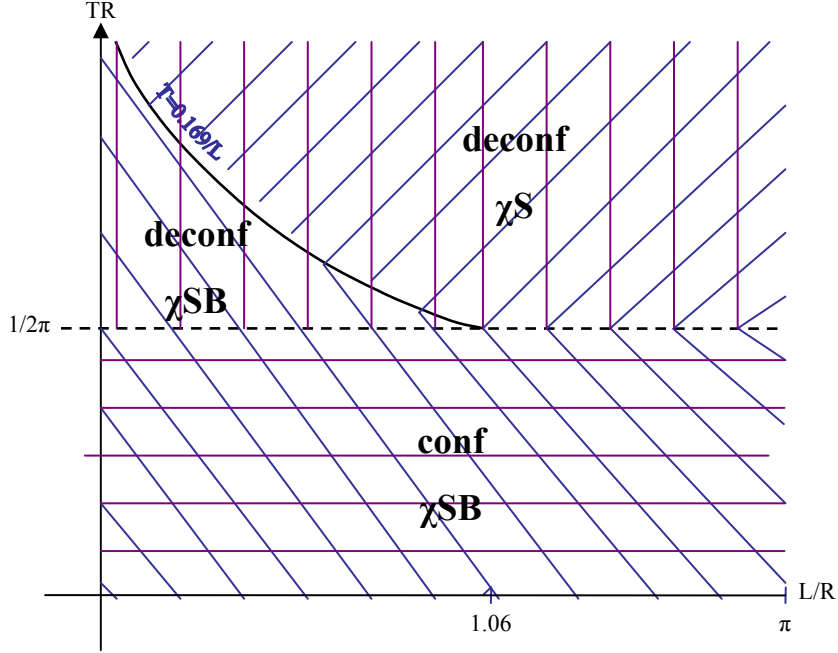}} \caption{The
phase diagram of the $AdS_6$ model with flavor D4-branes. The
phase structure depends only on the two dimensionless parameters
$TR$ and $L/R$. For $L/R<1.06$ the deconfinement and chiral
symmetry restoration transitions happens at different
temperatures, while for $L/R>1.06$ they occur together
.\label{phasediag''}}
\end{center}
\end{figure}

We would like to express the critical point in terms of physical
quantities. For a certain value of $y_T$ we can compute an
integral that relates the minimal point of the connected branes
configuration $u_0$ to the asymptotic distance between branes and
antibranes $L$.
\begin{align}\label{L}
L&=2\int_{u_0}^{\infty}\frac{du}{u'}=\frac{2R^2_{AdS}}{u_0}\int_1^{\infty}dy\frac{1}{y^2\sqrt{f(y)}}
\frac{1}{\sqrt{y^{10}\frac{f(y)}{f(1)}-1}}\nonumber\\
&
=\frac{2R_{AdS}^2}{5u_0}\sqrt{1-y_T^5}\int_0^1dz\frac{1}{\sqrt{1-y_T^5z}}
\frac{z^{\frac{1}{5}}}{\sqrt{1-y_T^5z-z^2(1-y_T^5)}}
\end{align}
For small values of $L$ $u_0\propto R^2_{AdS}/L$. At the
transition temperature $y_T^c=0.8$ the integral (\ref{L}) gives
$L=0.53(R^2_{AdS}/u_0)$. From the equation (\ref{beta}) we find
\begin{equation}\label{}
  T_{\chi SB}=\frac{5y_Tu_0}{4\pi R_{AdS}^2}=\frac{5y_T0.53}{4\pi
  L}=\frac{0.169}{L}
\end{equation}
while the deconfinement phase transition happens at the
temperature
\begin{equation}\label{}
  T_{d}=\frac{1}{2\pi R}=\frac{0.159}{R}
\end{equation}
Both temperatures are equal when $L=1.06R$. For $L/R>1.06$ and
$T\cdot R>T_d\cdot R$ the system is deconfined and chiral symmetry
is restored, while for $L/R<1.06$ and temperatures bigger than the
temperature of deconfinement the system is deconfined but chiral
symmetry restoration happens separately: at $T\cdot R<T_{\chi
SB}\cdot R$ chiral symmetry is still broken and at $T\cdot
R>T_{\chi SB}\cdot R$ chiral symmetry is restored. The full phase
diagram of the theory is drawn in figure \ref{phasediag''}.

\subsection{General model}

From the similarity of the result of the previous section to the
SS model we can derive a general model, but it is not necessary
that all three phases will be present. We indeed find a different
behavior of some metrics.

We consider a n-dimensional Wick rotated black hole background and
insert into it (n-2)-probe branes, that extend along all
directions except $x_4$. We take the following general form of the
metric at low temperatures:
\begin{equation}\label{dsgeneral'}
ds^2_n=H_2 dt^2+\frac{1}{H_1} du^2+H_1 dx_4^2+ds_k^2
\end{equation}
and then the metric at high temperatures becomes:
\begin{equation}\label{dsgeneral}
ds^2_n=H_1 dt^2+\frac{1}{H_1} du^2+H_2 dx_4^2+ds_k^2
\end{equation}
where $H_1$ is a singular function of u with a horizon at $u=u_H$,
$H_2$ is a non-singular function of $u$, $x_4$ is compact with a
period that depends on $u_H$ and $ds^2_k$ is any k-dimensional
metric of the rest of coordinates, whose components can depend on
$u$. The condition of the singular $H_1$ is necessary to have a
horizon on which (n-2)-branes can end. At low temperature the
$u-x_4$ submanifold is cigar-shaped, while at high temperature the
$u-t$ submanifold is cigar-shaped and the $x_4$ circle does not
shrink to zero.

From \eqref{dsgeneral} we find the induced metric on the
(n-2)-probe brane
\begin{equation}
ds^2_n=H_1 dt^2+(H_2+ \frac{1}{H_1}u'^2)dx_4^2+ds_k^2
\end{equation}
with $u'=du/dx_4$.

The DBI action is given by (without the CS term)
\begin{equation}\label{S_DBI}
S_{DBI}=T_n\int d^{n-1}xe^{-\phi}\sqrt{g}\sqrt{H_1(H_2+
\frac{1}{H_1}u'^2)}=\hat{T}_n\int
dx_4e^{-\phi}\sqrt{g}\sqrt{H_1(H_2+ \frac{1}{H_1}u'^2)}
\end{equation}
where $\hat{T}_n$ includes the outcome integration over all
coordinates apart from $dx_4$, $e^{\phi}$ is a dilaton and $g$ is
the determinant of the $ds^2_k$ metric. Then from the conservation
of the Hamiltonian we find that the equation of motion is
\begin{equation}\label{}
\frac{e^{-\phi}\sqrt{g}\sqrt{H_1}H_2}{\sqrt{H_2+
\frac{1}{H_1}u'^2}}=const
\end{equation}
If we suppose that there is a solution with $u'(u=u_0)=0$, then
the constant is equal to $e^{-\phi_0}\sqrt{g^0H_1^0H_2^0}$ and
\begin{equation}\label{}
u'=\sqrt{H_1H_2}\sqrt{\frac{e^{-2\phi}gH_1H_2}{e^{-2\phi_0}g^0H_1^0H_2^0}-1}
\end{equation}
The case $u'(u=0)\rightarrow\infty$ is also a solution of the
equation of motion and gives $const=0$.

Inserting the expression for $u'$ into \eqref{S_DBI} we find
\begin{equation}\label{}
S_{DBI}=2\hat{T}_n\int_{u_0}^\infty
due^{-\phi}\sqrt{g}\frac{1}{\sqrt{1-\frac{e^{-2\phi_0}g^0H_1^0H_2^0}{e^{-2\phi}gH_1H_2}}}
\end{equation}
At $u'\rightarrow\infty$ the actions reads
\begin{equation}\label{}
S_{DBI}^{u'\rightarrow\infty}=2\hat{T}_n\int_{u_0}^\infty
due^{-\phi}\sqrt{g}
\end{equation}
The action of the parallel brane configuration actually depends
only on the $x^k$ coordinates and the dilaton. The difference
between the actions is
\begin{equation}\label{deltaS_general}
\Delta S=2\hat{T}_n\int_{u_0}^\infty du
e^{-\phi}\sqrt{g}\left[\frac{1}{\sqrt{1-\frac{e^{-2\phi_0}g^0H_1^0H_2^0}{e^{-2\phi}gH_1H_2}}}
-1\right]-\int_{u_T}^{u_0} du e^{-\phi}\sqrt{g}
\end{equation}

To evaluate it we need to insert the explicit expressions of the
functions $H_1,H_2,g$ and the dilaton.

\begin{figure}[t]
\begin{center}
\scalebox{0.9}{\includegraphics[width=0.5\textwidth]{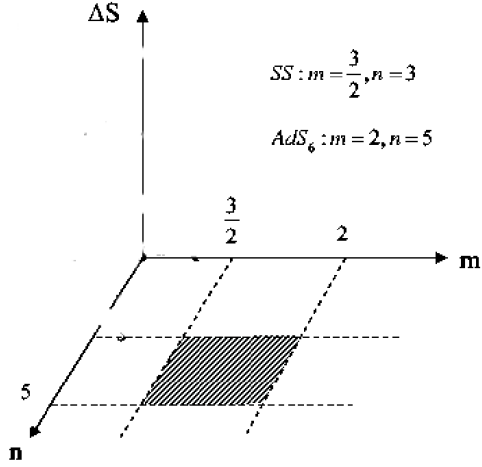}} \caption{Models
with values of $n$ between 3 and 5, $m$ between 3/2 and 2, and
arbitrary $j$ and $l$ should have intermediate and high
temperature phases.\label{n,m}}
\end{center}
\end{figure}

The existence of the solution with $u'(u=u_0)=0$ and $u_0\neq 0$
at low temperatures guarantees confinement, but to see whether we
get chiral symmetry restoration or not we need an explicit form of
the functions $\phi, g, H_1$ and $H_2$. Let us look at the same
family of metrics to which SS and $AdS_6$ metrics belong. That is:
\begin{align}\label{H1}
&
H_1=\left(\frac{u}{R}\right)^m\left(1-\left(\frac{u_T}{u}\right)^n\right)\nonumber\\
&H_2=\left(\frac{u}{R}\right)^m \nonumber\\
& e^{-\phi}=const_1\cdot u^l\nonumber\\
& g= const_2\cdot u^j
\end{align}
$const_1$ and $const_1$ does not effect $\delta S$ and therefore
we them arbitrary. Substituting the functions \eqref{H1} into the
action's difference \eqref{deltaS_general} and defining $y=u/u_0$
, we get:
\begin{equation}\label{DeltaS_general}
\Delta S\propto\int_1^{\infty}dy
y^{l+\frac{j}{2}}\left(\frac{1}{\sqrt{1-y^{-(2l+j+2m)}\frac{1-y_T^n}{1-(y_T/y)^n}}}-1\right)
-\int_{y_T}^1 dy y^{l+\frac{j}{2}}
\end{equation}

We did the numerical computations for different values of
$j,l,m,n$ (since we cannot solve the first integral analytically).
We checked that for $n$ between 3 and 5, for $m$ between 3/2 and
2, and arbitrary $j$ and $l$ $\Delta S$ is negative and then
positive as in the SS and $AdS_6$ models, as expected (see figure
\ref{n,m}).

\begin{figure}[t]
\begin{center}
\scalebox{0.9}{\includegraphics{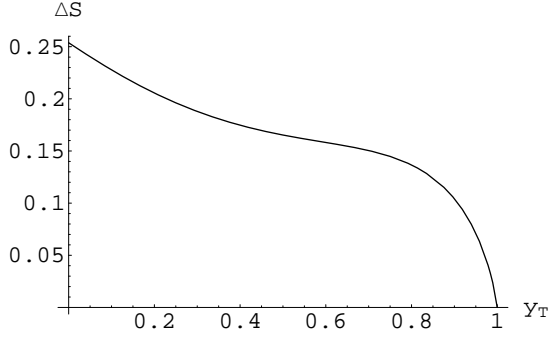}}
\caption{$\Delta S$ as a function of $y_T$ for the near extremal
D6-branes in 10 dimensions.\label{DeltaS p=6}}
\end{center}
\end{figure}

Also we can see whether we get a different phase structure for
general critical and non-critical versions of near extremal
Dp-branes. The metric for the critical near extremal Dp-branes is
\cite{Itzhaki:1998dd}:
\begin{align}\label{}
   ds^2&=\frac{u^{(7-p)/2}}{R_{Dp}^2}\left(-\left(1-
   \frac{u_T^{7-p}}{u^{7-p}}\right)dt^2+dx_4^2\right)+
   \frac{R_{Dp}^2}{u^{(7-p)/2}}\frac{1}{1-\frac{u_T^{7-p}}{u^{7-p}}}du^2\nonumber\\
   &+\frac{u^{(7-p)/2}}{R_{Dp}^2}\delta_{ij}dx^idx^j+R_{Dp}^2u^{(p-3)/2}d\Omega_{8-p}^2
   \qquad \quad i,j=1,...,p-1\nonumber\\
   e^{-\phi}&=\frac{1}{(2\pi)^{2-p}g_{YM}^2R_{Dp}^{3-p}}u^{(7-p)(3-p)/4}
   \quad \quad R_{Dp}^2=g_{YM}\sqrt{N}
\end{align}
Therefore all the powers depend only on $p$ and we find:
\begin{align}\label{}
  &m=\frac{7-p}{2}\nonumber\\
  &n=7-p\nonumber\\
  &l=\frac{(7-p)(3-p)}{4}\nonumber\\
  &j=\frac{(7-p)(p-1)+(p-3)}{2}
\end{align}
Since the solutions are in 10 dimensions, we insert D8-probe
branes because if there are directions along which probe branes do
not extend, except the $x_4$ direction, the massive quarks appear
and we will not get chiral symmetry. Drawing the $\Delta S$
\eqref{DeltaS_general} numerically for different values of $p$ we
find that for $p\leq5$ we get the same behavior as in the SS model
(two phases - chiral symmetry breaking and restoration), but for
$p=6$ we get a positive $\Delta S$ that means that chiral symmetry
restoration and deconfinement occur together (see figure
\ref{DeltaS p=6}). For $p>6$ we cannot get the solution of
\eqref{DeltaS_general} numerically.

Near extremal solutions of Dp-branes in non-critical dimensions
are given by \cite{Kuperstein:2004yk}:

\begin{align}\label{}
   ds^2&=\left(\frac{u}{R_{AdS}}\right)^2\left(-\left(1-
   \frac{u_T^{p+1}}{u^{p+1}}\right)dt^2+dx_4^2\right)+
   \left(\frac{R_{AdS}}{u}\right)^2\frac{1}{1-\frac{u_T^{p+1}}{u^{p+1}}}du^2\nonumber\\
   &+\left(\frac{u}{R_{AdS}}\right)^2\delta_{ij}dx^idx^j+R_{S^q}^2d\Omega_{q}^2
   \qquad \quad i,j \textrm{ arbitrary but smaller than }7-q\nonumber\\
   e^{-\phi_0}&=\left[\frac{1}{p+2-q}\left(\frac{(p+2-q)(q-1)}{c}\right)^q\frac{2c}{Q^2}\right]^{-1/2}
   \quad \quad R_{AdS}^2=\frac{(p+1)(p+2-q)}{c}
\end{align}
where
\begin{equation}\label{}
\frac{c}{\alpha'}=\frac{10-d}{\alpha'}
\end{equation}
is the non-criticality central charge term.

We see that again all the powers depend only on dimension of the
branes $p$:

\begin{align}\label{}
  &m=2\nonumber\\
  &n=p+1\nonumber\\
  &l=0\nonumber\\
  &j=6
\end{align}

The probe branes that we insert into the backgrounds are
(d-2)-branes and antibranes (d - is the dimension of a
non-critical metric). For different values of $n=p+1$ we find that
$\Delta S$ \eqref{DeltaS_general} always has the same behavior as
in the $AdS_6$ BH model, i.e. chiral symmetry can be restored at a
higher temperature then the temperature of deconfinement, but the
value of the ratio $L/R$ will be different for different
Dp-branes.


From the above analysis we see that the phase structure of a model
depends on the basic structure and dimensionality of its metric
and it can be different for different models.

%
%

\section{Spectrum of mesons}

The mesons of our model are described by strings ending on the
probe D4-branes. Low-spin mesons are described via modes of the
massless fields living on the D4-branes and high-spin mesons are
associated with string configurations that fall from the D4-branes
down to the wall at $u=u_{\Lambda}$, stretch along the wall and
then go back up again. The mesonic spectrum in the low-temperature
phase is unchanged as the temperature is increased because in the
confining phase the theory behaves effectively as a gas of
non-interacting glueballs and mesons
\cite{Neri:1983ic,Pisarski:1983db}. However, the mesonic spectrum
at intermediate temperature might be not connected to the spectrum
in the low-temperature regime since the phase transition is
first-order and such a jump should be expected. Now we turn our
attention to the meson spectrum at the intermediate and high
temperature phases.

\subsection{Low-spin mesons at intermediate temperature}

Low-spin mesons correspond on the string theory side to
fluctuations of the massless fields on the probe branes. The
fluctuations of the gauge fields on the branes give pseudo-vector
and scalar mesons and pions, and the fluctuations of the scalar
field describing the embedding of the branes give massive scalar
mesons. Using the analysis of the fluctuations performed in
\cite{Casero:2005se} we describe the modes coming from the
components of the gauge field living on the D4-branes.

The spectrum of low-spin mesons in the low-temperature phase is
unmodified with respect to zero temperature since the Euclidean
metric is globally unmodified.

The spectrum in the intermediate temperature phase is discrete
because the probe does not intersect the horizon. Also
computations in \cite{Peeters:2006iu} for the SS model show that
given that the effective tension of strings near the brane
decreases with the increase of temperature, the masses of mesons
decrease as the temperature is increased. This behavior is also
true for our model.

We start from the background metric describing the hot gluonic
plasma (\ref{ds6high}).The induced metric on the D4-brane
worldvolume at intermediate temperature reads
\begin{multline}
{\rm d}s_{\text{interm}}^2 = \left(\frac{u}{R_{AdS}}\right)^{2}
\left(f(u){\rm d}t^2 + \delta_{ij} {\rm d}x^i
{\rm d}x^j\right) \\[1ex]
+ \left[ \left(\frac{R_{AdS}}{u}\right)^{2} \frac{1}{f(u)}
         + \left(\frac{{\rm d}x^4}{{\rm d}u}\right)^2\, \left(\frac{u}{R_{AdS}}\right)^{2} \right]
  {\rm d}u^2
\end{multline}
We are interested in computing the spectrum of vector mesons, by
considering small fluctuations on the worldvolume gauge fields of
the probe D4-brane. We expand the gauge field as \cite{SS}
\begin{align}
&A_{\mu}(x^{\mu},u)=\sum_{n}B^{(n)}_{\mu}(x^{\mu})\psi_{(n)}(u)\\
&A_{u}(x^{\mu},u)=\sum_{n}\varphi^{(n)}_{\mu}(x^{\mu})\phi_{(n)}(u)
\end{align}
and therefore the field strength reads
\begin{equation}
\label{Fe}
\begin{aligned}
F_{\mu\nu} &= \sum_{n} F_{\mu\nu}^{(n)}(x^\rho)\,\psi_n(u)\,,\\[1ex]
F_{\mu u}  &= \sum_{n} \partial_\mu\varphi^{(n)}\,\phi_n(u) -
B^{(n)}_\mu \partial_u\psi_n(u) \\[1ex]
&= \partial_\mu \varphi^{(0)}\,\phi_{0}
  + \sum_{n\ge 1} \left( \partial_\mu \varphi^{(n)} - B^{(n)}_\mu \right)\partial_u \psi_{(n)}\,.
\end{aligned}
\end{equation}
where the last line is obtained by taking $\phi_{(n)} = m_n^{-1}
\partial_u \psi_{(n)}(u)$.  To simplify the consideration, we
furthermore go to the~$A_0=0$ gauge and consider only spatially
homogeneous modes, i.e.~we consider the equation of motion for
fields satisfying~$\partial_i A_j = 0$. Then the probe brane
action (with $a=0$ is
\begin{multline}
\hat{S}_{\text{trunc}} =
  \int\!{\rm d}^4x {\rm d}u\;
  u^4 \gamma^{1/2}\,f(u)^{1/2}
\, \bigg[ \frac{1}{u^2\gamma\,f(u)} (\partial_0\varphi^{(0)})^2\,\phi^{(0)}\phi^{(0)}\\[1ex]
-\frac{1}{f(u)}\left(\frac{R_{AdS}}{u}\right)^4\,
    \partial_0 B^{(m)}_i \partial_0 B_{(n)}^i\,\psi_{(m)}\psi_{(n)}
+ \frac{1}{u^2\gamma} B^{(m)}_i B_{(n)}^{i}\,\partial_u \psi_{(m)} \partial_u \psi_{(n)}\bigg]\,, \\
\text{with}\qquad \gamma \equiv  {\frac{u^8}{u^{10} f(u) -
u_0^{10}f(u_0)}} \,
\end{multline}
After a partial integration with respect to the $u$-coordinate,
the equation of motion for the field~$B_i^{(m)}$ becomes
\begin{eqnarray}
\frac{ u^2 }{\gamma^{1/2}f(u)^{1/2}}
\partial_0^2 B^{(n)}_i
\psi_{(n)} - \partial_u\left( u^2 \gamma^{-1/2} f(u)^{1/2}\, \,
\partial_u \psi_{(n)} \right) B^{(n)}_i = 0\, \,
\end{eqnarray}
This equation will reduce to the canonical form
\begin{equation}
\partial_0^2 B^{(n)}_i = - m_n^2 B^{(n)}_i \, ,
\end{equation}
if the modes $\psi_{(n)}$ satisfy the equation
\begin{equation}\label{midT}
- \gamma^{-1/2}\,f(u)^{1/2}\,\partial_u\left( u^2
 \gamma^{-1/2} f(u)^{1/2}\partial_u \psi_{(n)}\right) = R_{AdS}^4 \, m_n^2\, \psi_{(n)}\,.
\end{equation}
This equation is very similar to the equation in the zero
temperature case computed in \cite{Casero:2005se}, the only
difference is the appearance of the factor $f(u)^{1/2}$ in the
term on the left-hand side. The modes should also satisfy the
normalization conditions
\begin{equation}
\begin{aligned}
&\int_{u_0}^\infty \!{\rm d}u\, \gamma^{1/2} f(u)^{-1/2}
\,\psi_{(m)}\psi_{(n)} =
\delta_{mn}\,,\\[1ex]
&\int_{u_0}^\infty \!{\rm d}u\, \frac{u^2}{R_{AdS}^4}
\gamma^{-1/2} f(u)^{-1/2} \, \phi^{(0)} \phi^{(0)} = 1\,.
\end{aligned}
\end{equation}
The zero mode $\phi^{(0)} = u^{-2} f(u)^{-1/2} \gamma^{1/2}$ is
normalizable with this norm (there is no problem at the horizon
because $u_0 > u_T$), and therefore there is a massless pion
$\pi^{(0)}$ present in the intermediate-temperature phase. The
fields $\pi^{(0)}$ are the Goldstone bosons associated with the
spontaneous breaking of the $U(N_f)_L\times U(N_f)_R$ global
chiral symmetry to the diagonal $U(N_f)$.

In the limit of $u_0 \gg u_T$ the spectrum simplifies and one can
easily determine the scale of the meson masses.  In this limit,
which corresponds to a small separation distance between the
stacks of branes and anti-branes $L \ll R$, the thermal factor
$f(u)\rightarrow 1$ and in particular also $f(u_0)\rightarrow
1$.Therefore,
\begin{equation}
\gamma\equiv\frac{u^8}{u^{10} f(u) -
 u_0^{10} f(u_0)}\rightarrow \frac{1}{u^2}\frac{1}{1-y^{-10}}
\end{equation}
where the dimensionless quantity $y\equiv u/u_0$. Then we can
rewrite~\eqref{midT} in terms of~$y$ in the following form
\begin{equation}
- \gamma^{-1/2}(y) \partial_y \left(
   y^2 \gamma^{-1/2}(y) \partial_y \psi_{(n)}
\right) = \frac{R_{AdS}^{4}}{u_0^2}\,m_{n}^2\,\psi_{(n)} \,.
\end{equation}
Now since the left-hand side is expressed in terms of the
dimensionless quantity~$y$, the right-hand side should also be
dimensionless which implies that
\begin{equation}
\label{scaling} m_n^2 \sim \frac{u_0^2}{R_{AdS}^4}
\end{equation}
From (\ref{L}) we know that $u_0\sim 1/L$. Therefore the mass of
``short'' mesons scales as
\begin{equation}
M_{\text{meson}}\sim \frac{1}{L}\,.
\end{equation}
The explicit mass spectrum of the vector mesons can be found by
looking for normalizable eigenfunctions of (\ref{midT}) and using
numerical methods (e.g. a shooting technique), but the qualitative
behavior of the spectrum is that the masses of mesons decrease as
temperature increases. This behavior is a direct consequence of
the fact that the constituent quark mass is related to the
distance of the tip of the probe brane to the horizon. If the
distance is increased, a meson of the same spin will correspond to
an excitation of the brane which is further away from the horizon
and hence less affected by the temperature. This behavior is
common for all gravitational backgrounds that contain a horizon.

\subsection{Low-spin mesons at high temperature}

In the high-temperature phase the profile of the left and right
stacks of branes is characterized by $u'\equiv {\rm d}u/{\rm
d}x^4\rightarrow \infty$ and the induced metric on the probe
branes and probe anti-branes takes the form
\begin{equation}
\label{e:highTmetric} {\rm d}\hat{s}^2_{\text{high}}=\left(
\frac{u}{R_{AdS}} \right)^{2}
  \left[ -f(u) {\rm d}t^2+  \delta_{ij}{\rm d}x^{i}{\rm d}x^j \right ]
+\left( \frac{R_{AdS}}{u} \right)^{2}\frac{1}{f(u)} {\rm d}u^2
\end{equation}
The differential equation for the modes is now
\begin{equation}
\label{e:eqhighT} -\,f(u)^{1/2}\,\partial_u\left( u^{2}
 f(u)^{1/2}\partial_u \psi_{(n)}\right) = R_{AdS}^4 \, m_n^2\, \psi_{(n)}
\end{equation}
i.e. it is similar to the intermediate temperature phase case, but
with $\gamma=1$. Then the normalization conditions are now
\begin{equation}
\begin{aligned}
&\int_{u_0}^\infty \!{\rm d}u\, f(u)^{-1/2} \,\psi_{(m)}\psi_{(n)}
=\delta_{mn}\,,\\[1ex]
&\int_{u_0}^\infty \!{\rm d}u\, \frac{u^2}{R_{AdS}^4} f(u)^{-1/2}
\, \phi^{(0)} \phi^{(0)} = 1\,.
\end{aligned}
\end{equation}

The mode, which would be given by~$\phi^{(0)} = u^{-2}
f(u)^{-1/2}$, is no longer normalizable. Computation of its norm
leads to the integral
\begin{equation}
\int_{u_T}^\infty\!{\rm d}u\, u^{2} f(u)^{-1/2} \Big| u^{-2}
f(u)^{-1/2}\Big|^2\,,
\end{equation}
which, while convergent at the upper boundary, is divergent at the
lower boundary because~$f(u) \sim \sqrt{u-u_T}$ for $u\sim u_T$.
In accordance with the fact that chiral symmetry is restored in
the high-temperature phase, we see that the Goldstone boson has
disappeared. In the high temperature phase the spectrum of vector
mesons is continuous.

\subsection{High-spin mesons at intermediate temperature}

\begin{figure}[t]
\begin{center}
\includegraphics[width=.9\columnwidth]{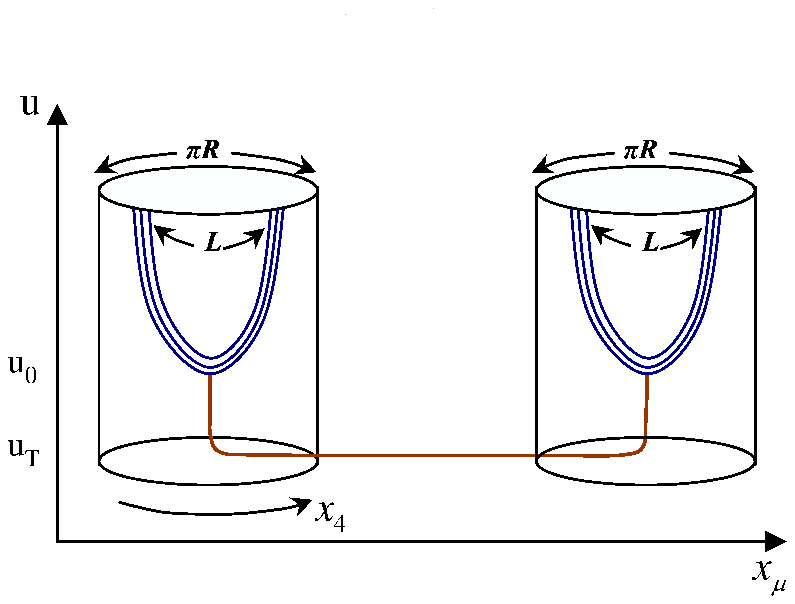} \caption{A high-spin
meson at intermediate temperatures represented as a semiclassical
string starting at the lowest point of the probe brane $u=u_0$,
going down to the wall at $u=u_T$, stretching horizontally in the
space along the wall, and then going back up vertically to the
probe brane at $u=u_0$.\label{classstring}}
\end{center}
\end{figure}

To describe higher-spin mesons we cannot use supergravity modes and
we need to consider
string configurations \cite{Kruczenski:2004me} that start and end on  probe branes. For
large spin these strings can be described semiclassically. The
relevant string configuration can be decomposed into three parts:
a segment from the probe brane at $u=u_0$ to the wall at $u=u_T$,
then a segment that stretches along the wall in the spacial
direction, and then another vertical part stretching from the wall
back to the probe brane, as depicted at figure \ref{classstring}.

The relevant part of the background metric that represents this
configuration is
\begin{equation}
{\rm d}s^2 = \left(\frac{u}{R_{AdS}}\right)^{2} \left( - f(u)\,
{\rm d}t^2 + {\rm d} \rho^2 + \rho^2\,{\rm d} \varphi^2 \right) +
\left(\frac{R_{AdS}}{u}\right)^{2} \frac{{\rm d}u^2}{f(u)}\,
\end{equation}
We go to the static gauge for the string action and make the
following ansatz for the rotating configuration,
\begin{equation}
\label{ansatz-meson} t = \tau \, , \quad \rho=\rho(\sigma)\,,\quad
u = u(\sigma)\,,
 \quad \varphi = \omega \tau
\end{equation}
This ansatz has the same form as in the zero-temperature case
\cite{Kruczenski:2004me}. Hence, the only effect of finite
temperature will be in the change of the shape of $u(\sigma)$ as
the temperature is increased.

With this ansatz the metric now reads
\begin{equation}
{\rm d}s^2 = \left(\frac{u}{R_{AdS}}\right)^{2} ( - f(u)\, {\rm }
+  \rho^2\omega^2){\rm d} \tau^2  +
\bigg(\frac{u}{R_{AdS}}\bigg)^{2}\bigg(
\rho'^2+\bigg(\frac{R_{AdS}}{u}\bigg)^4\frac{u'^2}{f(u)}\bigg)d\sigma^2
\end{equation}
and it leads to the following  string (Polyakov) action
\begin{align}\label{Sind}
 S = \int\!{\rm d}\tau \,  {\rm d}\rho \,
\sqrt{\left(\frac{u}{R_{AdS}}\right)^4\,\left(\rho'^2  +
\frac{u'^2}{f(u)} \frac{R_{AdS}^4}{u^4}\right) \left( f(u) -
\rho^2\omega^2 \right)}
\end{align}
Positivity of the argument of the square root in \eqref{Sind}
requires that \mbox{$f(u)> \rho^2 \omega^2$}. This means that for
a given angular frequency $\omega$, the string solution $u(\rho)$
has to lie above the curve
\begin{equation}\label{u>}
u(\rho) \geq \frac{u_T}{(1 - \rho^2 \omega^2)^{1/5}}
\end{equation}

\begin{figure}[ht*]
\begin{center}
\includegraphics[width=.5\textwidth]{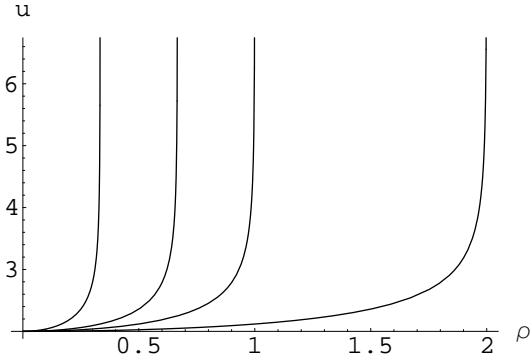}
\end{center}
\caption{The boundary curve. Rotating strings
  have to lie above this curve in order for their action to be
  real. The curves correspond to the following values of the
  frequency~$\omega$ (from right to left): 0.5, 1, 1.5 and 3. The
  horizon is located at $u_T=2$.
\label{u(p)}}
\end{figure}

In figure \ref{u(p)} these curves are depicted for various values
of $\omega$. We see that any string is allowed to touch the
horizon $u_T$ for any angular frequency $\omega$ and as $\omega$
decreases (i.e. the spin of the mesons increases) the string
endpoints get more and more separated, the U-shaped string
penetrates deeper to the horizon, and it becomes more and more
rectangular. For a given~$\omega$, the maximal allowed extent of
the string is determined by the intersection of the curve
with~$u_0$, and is given by
\begin{equation}
\rho_{\text{max}} = \frac{1}{\omega}\sqrt{1 -
\bigg(\frac{u_T}{u_0}\bigg)^5 }\,
\end{equation}

The equation of motion following from the action (\ref{Sind}) is
given by
\begin{multline}
\label{equ}
 -2 \sqrt{...} \frac{{\rm d}}{{\rm d}\sigma} \left( \frac{1}{\sqrt{...}} \frac{u'}{f(u)} (f(u) - \rho^2 \omega^2) \right)
+ f'(u)\bigg(\frac{u'^2 \rho^2 \omega^2}{f(u)^2} +
\frac{u^4}{R_{AdS}^4}
(\rho')^2 \bigg) \\[1ex]
+ \frac{4 u^3}{R_{AdS}^4} \bigg((\rho')^2f(u)  -(\rho')^2 \rho^2
\omega^2 \bigg) = 0
\end{multline}

where $\sqrt{...}$ is the density of Nambu-Goto
action~\eqref{Sind}.

The expressions for the energy and the angular momentum carried by
the string are given by
\begin{align}
\label{e:Edef} E &= \int\!{\rm d}\sigma\, \frac{1}{\sqrt{...}}
\bigg(  \left(\frac{u}{R_{AdS}}\right)^4 f(u)
(\rho')^2  + u'^2 \bigg) \\[1ex]
\label{e:Jdef} J &= \int\!{\rm d}\sigma\, \frac{1}{\sqrt{...}}
\omega \rho^2 \bigg(  \left( \frac{u}{R_{AdS}} \right)^4\,\rho'^2
+ \frac{u'^2}{f(u)} \bigg)
\end{align}

The analysis of meson spectrum in \cite{Peeters:2006iu} has showed
that the meson spectrum of the SS model does not follow the well
known Regge trajectories. For high-spin mesons at a fixed
temperature there is a maximum value of angular momentum beyond
which mesons cannot exist and have to dissociate. That is the
temperature at which mesons melt is spin dependent. As the
temperature increases, the maximal value of the spin that a meson
can carry decreases. This behavior is also true for high-spin
mesons in the $AdS_6$ background. Also for high mesons of fixed
angular momentum, as for low-spin mesons, the energy decreases as
a function of temperature.

\subsection{Drag effects for quarks}

In the deconfined phase the background contains a horizon and we
can have a string starting on a flavor D4-brane/antibrane and
going into the horizon. This string corresponds to a deconfined
quark/anti-quark.

Because a strictly vertical string moving rigidly through the
background would not have a real action \eqref{Sind}, the string
has to be bent when it is ``pushed'' through the plasma. In
addition the bent string does not end anymore orthogonally on the
brane. This means that one has to apply a force on the string
endpoint, or in other words, one has to ``drag'' the string in
order to keep it moving
\cite{Herzog:2006gh,Casalderrey-Solana:2006rq,Gubser:2006bz,Herzog:2006se,
Caceres:2006dj,Friess:2006aw,Sin:2006yz}. A suitable ansatz to
describe the behavior of the string that moves with speed $v_x$ in
the $x$ direction is (in static gauge)
\begin{equation}
\label{ansatz-quark} t=\tau \,, \quad u= \sigma\,, \quad x = v_x
t+\xi(u)
\end{equation}

Inserting (\ref{ansatz-quark}) into the Nambu-Goto Lagrangian we
find
\begin{align}\label{actionNG}
 S &=\int d^2\sigma\sqrt{-det(G_{\mu\nu}\partial_{\alpha}X^{\mu}\partial_{\beta}X^{\nu}) }\nonumber\\
 &= \int\!{\rm d}\tau \,  {\rm d}\rho \,
\sqrt{1 -
\frac{v_x^2}{f(u)}+\left(\frac{u}{R_{AdS}}\right)^4f(u)\xi'^2}
\end{align}
The corresponding equation for $\xi$ implies that the conjugate
momentum is a constant:
\begin{equation}
\pi_{\xi}=\frac{\partial
\mathcal{L}}{\partial\xi'}=-\left(\frac{u}{R_{AdS}}\right)^4\frac{f(u)\xi'}{\sqrt{-g}}
\end{equation}
where $g$ is the determinant of the induced metric. Inverting this
relation we obtain
\begin{equation}\label{xi}
\xi'=\pi_{\xi}\bigg(\frac{R_{AdS}}{u}\bigg)^4
\frac{1}{f(u)}\sqrt{\frac{f(u)-v_x^2}{f(u)-\pi^2_{\xi}(\frac{R_{AdS}}{u})^4}}
\end{equation}
We must require that $\xi(u)$ is everywhere real, but the square
root on the right hand side is in general not everywhere real. The
function $f(u)$ interpolates between 1 at the boundary of $AdS_6$
to 0 at the horizon, so at some intermediate radius $f(u)-v_x^2$
switches sign at some intermediate point $u_v$, which is by
definition such that $u_v^5=u_T^5/(1-v_x^2)$. The only way we can
prevent $\xi$ from becoming imaginary for $u<u_v$ is by choosing a
value of $\pi_{\xi}$ such that the denominator also vanishes at
$u_v$:
\begin{equation}
\pi_{\xi}^2=f(u_v)\left(\frac{u_v}{R_{AdS}}\right)^4=
\left(\frac{u_T}{R_{AdS}}\right)^4\frac{v_x}{(1-v_x^2)^{\frac{4}{5}}}
\end{equation}
Plugging this back into (\ref{xi}) we find
\begin{equation}\label{xi'}
 \xi'=\frac{vR_{AdS}^2}{u^4}\frac{u_T^2}{f(u)}
\end{equation}
Now we want to compute the $\sigma$ component of the current
associated with spacetime translations along $x$
\begin{equation}
 P_x^u=-G_{x\nu}g^{u\alpha}\partial X^{\nu}
 =-\frac{f(u)\xi'}{g}\left(\frac{u_T}{R_{AdS}}\right)^4
\end{equation}
where $G_{\mu\nu}$ and $g_{\alpha\beta}$  denote respectively the
spacetime and induced worldsheet metric. Together with
(\ref{actionNG}) and (\ref{xi'}) it yields the drag force
\begin{equation}\label{}
\frac{dp}{dt}=\sqrt{-g}P_x^u=-\frac{u_T^2}{R^2}\frac{v_x^2}{(1-v_x^2)^{\frac{4}{5}}}
\end{equation}
We see that there are two effects happening as one tries to move a
single string in the hot background: the string shape is modified
in a way which depends on the temperature and velocity, and in
order to preserve the motion one needs to apply a force.

\subsection{Drag effects for mesons}

Now we are interested if there is a drag force on a rotating meson
at finite temperature. From the condition (\ref{u>}) we can see
that a simple rotating motion does not experience a drag effect
because the rotating string is always sufficiently high above the
curve beyond which the action would turn to be imaginary. On the
other hand the bending of the rotating string does depend on the
angular velocity and on the temperature.

We can also consider a linear motion of the meson in a direction
orthogonal to the plane of rotation. A suitable ansatz for this
motion is
\begin{equation}
\label{e:movingmeson}
 t =\tau \, , \quad  \rho = \sigma \, , \quad u=u(\rho) \, , \quad
 \varphi  = \omega \tau  \, ,
 \quad  y = v_y\, \tau
\end{equation}
In this case the string action becomes
\begin{equation}
\label{Sindv} S = \int\!{\rm d}\tau \,  {\rm d}\rho \,
\sqrt{\left(\frac{u}{R_{AdS}}\right)^4\,\left(1  +
\frac{u'^2}{f(u)}
  \frac{R_{AdS}^4}{u^4}\right) \left( f(u)  - \rho^2\omega^2 - v^2_y\right)} \,  .
\end{equation}
The only modification with respect to the rotating meson is the
addition of a term~``$-v_y^2$'' to the last factor under the
square root. The condition for the action to be real is now
\begin{equation}
\label{e:minuv} u \geq \frac{u_T}{(1-\rho^2 \omega^2
-v_y^2)^{1/5}}\,.
\end{equation}

\begin{figure}[t]
\begin{center}
\scalebox{1.2}{\includegraphics[width=.48\textwidth]{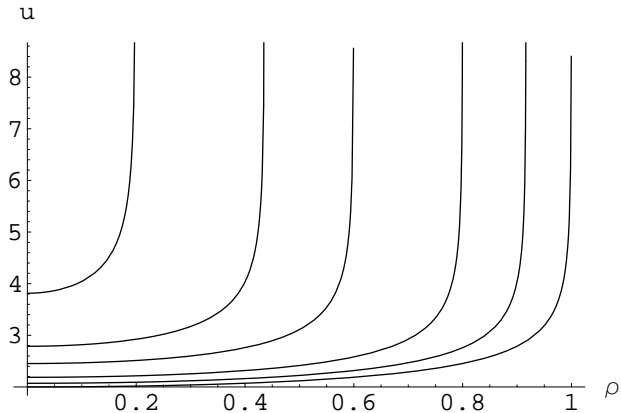}}
\end{center}
\caption{Analysis of the effect of a transverse velocity on the
shape of spinning U-shaped strings keeping the quark masses and
spin fixed. The curves display results for increasing values (from
right to left)~$v_y=0, 0.4, 0.6, 0.8, 0.9, 0.98$ and $\omega=1$.
The horizon is located at $u_T=2$. \label{u(p)2}}
\end{figure}

The curves are depicted in figure \ref{u(p)2} for various values
of $v_y$. For any high-spin meson at finite temperature one can
find a generalized solution to the equation of motion such that
the spinning configuration lies entirely above the curves
\eqref{u(p)2}. Thus, the mesons do not experience any drag effect
that means that they do not experience any energy loss when
propagating through the quark-gluon plasma - no force is necessary
to keep them moving with a fixed velocity. In the dual language,
this reflects the fact that if the quark gluon plasma is not hot
enough to dissociate mesons, then these color singlets will not
experience a drag force generated by a monopole interactions with
the medium. However, the shape of the string in the ($\rho,u$)
plane can be modified as it starts moving.

\section{Summary}

In this project  we were mainly interested in the difference between
HQCD
models in critical and non-critical backgrounds. We looked at the
thermal properties of the
non-critical $AdS_6$ black hole background with D4-probe branes
and antibranes. We   found that for the chargeless flavor branes, namely,
when we switched off the $\int c_5$ CS term,
 it has a similar behavior to  the
critical Sakai-Sugimoto model.

We found that the difference in free energy between Euclidean
background at low and high temperatures scales as $N_c^2$ in the 't
Hooft large $N_c$ limit, as expected. In our model it is
proportional to $(2\pi T)^5-(1/R)^5$, while in the SS model it is
proportional to $(2\pi T)^6-(1/R)^6$.
It seems that the former result fits better a five dimensional field theory.
 In our model we got the chiral phase transition at the ratio of
the separation distance between branes and antibranes at infinity
and the radius of the $x_4$ circle $L/R=1.06$, comparatively to
the SS model $L/R=0.97$. Spectrum of the low-spin mesons is
discrete at low temperatures and continuous at high temperatures
and we can identify the Goldstone pion associated with the
spontaneous breaking of the $U(N_f)_L\times U(N_f)_R$ global
chiral symmetry to the diagonal $U(N_f)$. Quarks and high-spin
mesons experience a drag force at finite temperature whereas mesons
do not. This is the same as was discovered in \cite{Peeters:2006iu}.

 We saw that the $AdS_6$ and the SS model can be
unified into a family of metrics assoicated with space-times
of different dimensions  that have a similar phase
structure.

The non-critical has a serious drawback and that is the fact that
the background has curvature of order one.
This property implies that higher curvature corrections may be important.
However, the resemblance with the critical picture indicates that presumably
for the properties considered in this work the higher curvature correction are not very
meaningful. On the other hand in the critical case
the  dilaton
goes to infinity as $u\rightarrow\infty$, thus, in principle,
in this region the sugra approximation is not valid and one should go to the
M-theory.
In the $AdS_6$ case dilaton is constant and we should
not worry about the behavior of the model when
$u\rightarrow\infty$. Again inspite of this difference the extracted physical properties
seem to be alike.
\section*{Acknowledgements}
We would like to thank Ofer Aharony and Stanislav Kuperstein for useful conversations.
This work was  supported in part by a center of excellence
supported by the Israel Science Foundation (grant number 1468/06), by
a grant (DIP H52) of the German Israel Project Cooperation, and by the
European Network MRTN-CT-2004-512194


\newpage{}


\end{document}